# Ultra-wideband radar to measure *in vivo* musculoskeletal forces

Christopher S. Bird[1,2], Antonio P. L. Bo[3,4], Wei Lu[2,4], Taylor J. M. Dick[1*]

[1]School of Biomedical Sciences, The University of Queensland, St Lucia, Australia
[2]Australian e-Health Research Centre, Commonwealth Scientific and Industrial Research Organisation, Herston, Australia
[3]Department of Mechanical, Electrical and Chemical Engineering, Oslo Metropolitan University, Oslo, Norway
[4]School of Electrical Engineering and Computer Science, The University of Queensland, St Lucia, Australia

*Author for correspondence: t.dick@uq.edu.au

# Abstract

Accurate measures of musculoskeletal forces are critical for clinicians, biomechanists, and engineers, yet direct measurement is highly invasive and current estimation methods remain limited in accuracy. Here, we demonstrate the application of ultra-wideband radar to non-invasively estimate musculoskeletal forces by measuring changes in the electromagnetic properties of contracting muscles, in muscles with different structural properties, during various static and dynamic conditions, and in the presence of fatigue. First, we show that ultra-wideband radar scans of muscle can reliably track isometric force in a unipennate knee extensor (vastus lateralis) and a bipennate ankle dorsiflexor (tibialis anterior). Next, we integrate radar signals within machine-learning and linear models to estimate musculoskeletal forces during fatiguing isometric, and dynamic knee extension contractions, with exceptional accuracy (test $R^2$>0.984; errors<3.3%). Finally, we identify frequency-dependent effects of musculoskeletal forces on ultra-wideband radar signals, that are independent of physiological and structural features known to influence muscle force. Together, these findings establish ultra-wideband radar as a powerful, non-invasive approach for quantifying *in vivo* musculoskeletal forces, with transformative potential for wearable assistive technologies, biomechanics, and rehabilitation.

# Main

Skeletal muscles are the engines of motion – producing forces to enable locomotion and interaction with the world. The quest for a tool capable of non-invasively measuring musculoskeletal forces has eluded clinicians, biomechanists, and engineers for decades. Accurate values of muscle force have application in biomechanics for developing and validating robust muscle models or musculoskeletal simulations; in sports science for risk assessment and injury prevention; and in engineering for real-time control of wearable assistive robotic devices to enhance or restore motor performance. Despite this wide range of applications, accurately quantifying *in vivo* muscle forces remains a major challenge. Direct measurements with implanted tendon transducers are highly invasive [1, 2, 3, 4], while the indirect estimates that are more commonly used are constrained by several methodological and technical challenges.

A range of methodologies have been developed to quantify musculoskeletal forces – each with important limitations. Electromyography (EMG) measures the electrical activation of muscle and can approximate force in highly constrained contractions of a few muscles [5], however its estimates become unreliable under conditions such as muscle fatigue [5, 6, 7, 8, 9]. Shear-wave elastography infers muscle forces based on estimates of tissue stiffness; but low sampling rates limit its application for real-time measures or dynamic movements [10, 11, 12]. Shear-wave tensiometry improves temporal resolution by tracking wave propagation in tendons, but validation has been restricted to superficial tendons, and accuracy varies with tendon morphology and loading conditions [13, 14, 15, 16, 17]. Alternatively, joint torques can be directly measured with a load cell or dynamometer. However, this limits the range and types of motions that can be measured, and as these are measures of net joint torque, load-sharing between agonist muscles or co-activation of antagonist muscles make it difficult to determine individual muscle forces [18]. Collectively, these limitations highlight the need for new methods capable of providing accurate and generalizable measurements of musculoskeletal forces across muscles, movements, and loading conditions.

Ultra-wideband (UWB) radar consists of low energy, non-ionising electromagnetic radiation across a wider frequency band than conventional radar, providing information across a range of frequencies. As the electromagnetic properties of a material changes, so too does the propagation of electromagnetic waves through that material, which can be measured from the UWB radar signal. The differences in electromagnetic properties of various human tissues are utilised in electromagnetic imaging [19, 20, 21] and beyond imaging, for example non-invasive estimates of blood glucose concentration [22, 23, 24], whereby the change in electromagnetic properties of blood can be characterised at different concentrations of glucose.

It is well established that the electrical impedance of a muscle, which is closely related to its electromagnetic properties, changes from a rested to contracted state [25, 26], and this change can be related to force magnitude during constrained isometric contractions [27, 28, 29, 30]. This is the theory behind electrical impedance myography techniques that have emerged recently [27, 30, 31, 32], which aim to provide a window into muscle behaviour by tracking impedance changes during contraction. During skeletal muscle contractions, the microscopic actin-myosin machinery is modulated via intracellular $Ca^{2+}$ [33, 34], which may be responsible for some of the measured changes in impedance. However, the accuracy of electrical impedance-based estimates of musculoskeletal forces remains limited [31, 32], potentially due to electrical impedance measures of skeletal muscle also being impacted by neuromuscular changes associated with muscle fatigue [27, 35] and structural changes such as muscle fibre orientation [25, 36, 37, 38]. Numerous factors influence a muscle's active force production, including its level of activation, fatigue status, and fibre/fascicle dynamics (i.e., operating length and velocity) [39, 40]; any tool capable of accurately measuring muscle forces must provide reliable estimates despite variations in these neural, structural, and physiological properties. Fortunately, the electromagnetic properties of body tissues are frequency dependent [41, 42]. If the changes in these properties caused by different neuromuscular, structural, or physiological factors are linearly independent across a range of frequencies, then sampling across these frequencies with UWB radar may allow for such structural changes to be accounted for when quantifying musculoskeletal forces.

In this work, we developed and tested ultra-wideband (UWB) radar as a tool to non-invasively estimate *in vivo* musculoskeletal forces, by measuring changes in the electromagnetic properties of contracting skeletal muscles. Our goal was to assess the capacity of UWB radar to measure skeletal muscle forces in a range of contractile conditions from isometric to dynamic, across different muscle architectures, and in the presence of fatigue. In a series of experiments, we combined UWB radar, surface electromyography (sEMG), b-mode ultrasound, and dynamometry to determine how these neuromuscular and structural properties influence UWB-radar signals. First, we determined how UWB radar scans of isometrically contracting skeletal muscle are impacted by changing levels of muscle force, in two muscles with different structural properties – the vastus lateralis (VL), a unipennate knee extensor and the largest of the quadriceps; and the tibialis anterior (TA), a bipennate ankle dorsiflexor. Next, we used UWB radar scans of the VL during a fatiguing set of intermittent isometric knee extension contractions, to train long short-term memory (LSTM) machine-learning models that estimate skeletal muscle forces in the presence of varying levels of muscle fatigue. Finally, we used linear models to estimate knee torques from UWB radar scans of the VL across various dynamic knee extension contractions (passive, isokinetic, and isotonic), where forces were decoupled from activation and fascicle length.

# Results

## Ultra-Wideband Radar

Ultra-wideband (UWB) radar data were collected using one and two body-coupled antennas [43] (Fig. 1a). A single antenna allows the S11 (input reflection coefficient) to be captured, measuring the signal reflected back to a transmitting antenna (Fig. 1b). With a second antenna, the S21 (forward transmission coefficient) can be captured, which measures the signal received by the second antenna that was transmitted by the first (Fig. 1b). These scattering parameters were found for each of 51 sampled frequencies from 100 MHz to 2.9 GHz using a vector network analyser (VNA), with changes in both magnitude and phase measured. We performed four different experiments to explore the use of UWB radar scans as a tool to estimate forces in muscles with different architectures and across various contraction conditions (Fig. 1c).

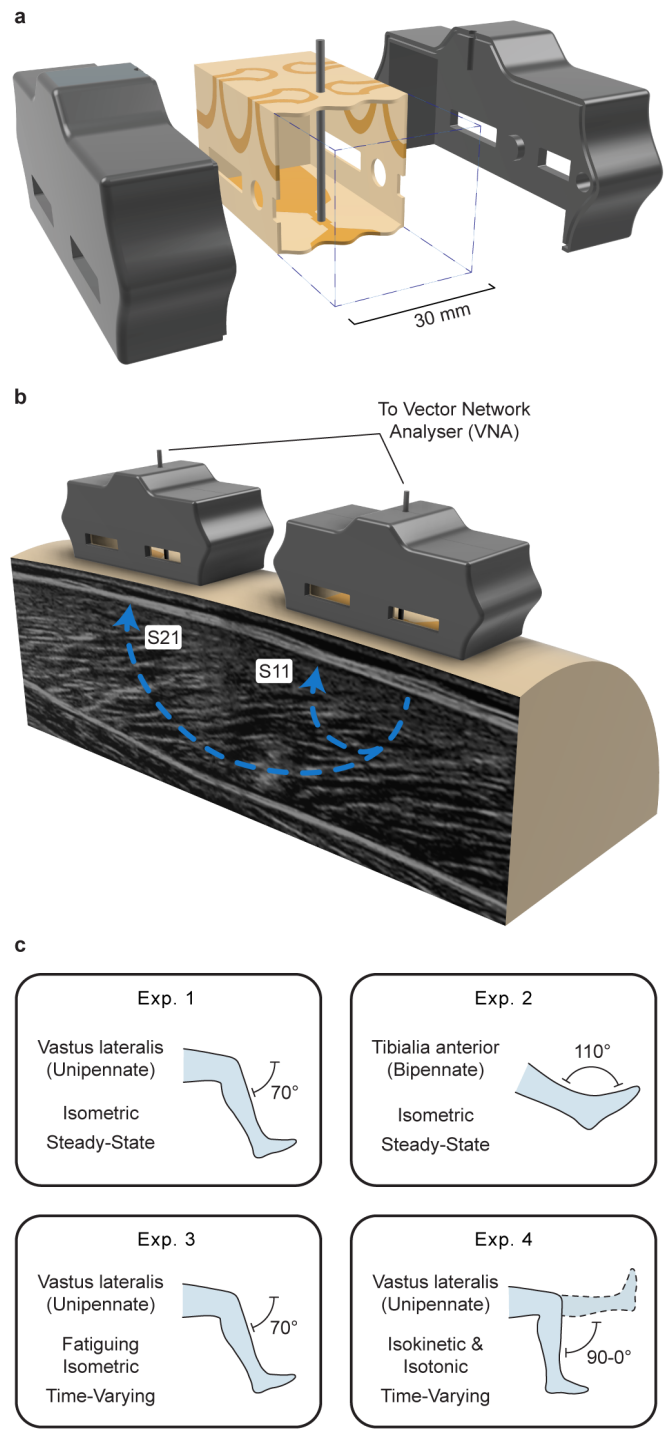

**Figure 1: Experimental Overview.**
**a**, Body-coupled UWB antenna (30×30×70 mm) with 3D printed housing. **b**, Representation of reflection coefficient (S11) and forward transmission coefficient (S21) signal paths. **c**, Overview of the four experiments, across muscle architectures and joint/contraction conditions.

## Isometric Knee Extension Contractions

Our investigation into the effect of muscle forces on UWB radar scans started with the simplest condition: varying levels of steady-state force during isometric contractions, where we expected changes in force to be correlated with changes in activation and fascicle length . For this condition, we quantified the absolute change in the magnitude and phase of S11, averaged across all sampled frequencies. While these were simplified measures that removed the complex frequency-dependence of the signals, we still expected that changes in the signal related to muscle forces would be preserved under these controlled conditions. Later experiments under more complex conditions made use of the full UWB radar signal.

We performed a study on 16 participants, measuring signals in the unipennate VL during isometric knee extension contractions at 10, 20, 30, 40, and 50% of their maximum voluntary contraction (MVC) with biofeedback. We used a VNA, sEMG, and b-mode ultrasound to measure UWB radar S11, activation, and fascicle length in the VL, respectively, and a load cell to measure knee extension torques (Fig. 2a). Here, we use knee extension torque and VL force (when normalised to MVC) synonymously, assuming minimal changes in load-sharing between VL and vastus medialis (VM) and limited changes in moment arm during the moderate-intensity fixed-end contractions. We found a significant fixed effect of VL force on the change in S11 magnitude ($p<0.001$, Fig. 2b) and S11 phase ($p<0.05$, Fig. 2c). As expected, we also found a significant fixed effect of VL force on both activation ($p<0.001$, Fig. 2d) and fascicle length ($p<0.001$, Fig. 2e), with both remaining well correlated with force (both $R^2>0.9$) during the isometric contractions.

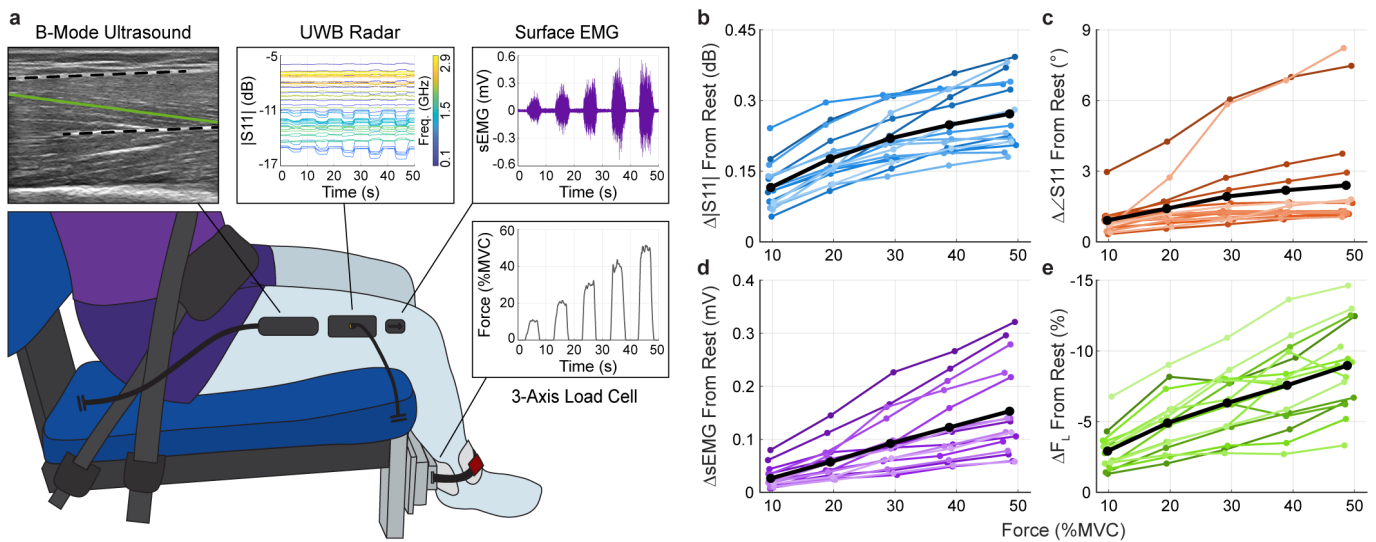

**Figure 2: VL forces during steady-state isometric knee extension contractions had a significant fixed effect on UWB radar S11 scans in both magnitude and phase.**
**a**, Illustration of isometric knee extension contractions experimental setup. **b**, Absolute change in S11 magnitude in the VL from rest, averaged across all frequencies (n = 16, $p<0.001$, LME). **c**, Absolute change in S11 phase in the VL from rest, averaged across all frequencies (n = 16, $p<0.05$, LME). **d**, Absolute change in sEMG magnitude in the VL from rest (n = 16, $p<0.001$, LME). **e**, Absolute change in fascicle strain in the VL from rest (n = 16, $p<0.001$, LME). For **b-e**, data is displayed for each participant (by lightness) with mean (black).

## Isometric Ankle Dorsiflexion Contractions

To explore whether the relationship between UWB radar and muscle force exists across muscles of varying architectures, we investigated the bipennate TA – an ankle dorsiflexor. We performed a study on 6 participants during isometric ankle dorsiflexion contractions at 10, 20, 30, 40, and 50% of MVC with biofeedback, analysing the same simplified UWB radar measures as the isometric knee extension contractions, against steady-state isometric force level. We combined a VNA, sEMG, and b-mode ultrasound to measure UWB radar S11, activation, and fascicle length of the TA, with a torque sensor used to measure ankle dorsiflexion torques (Fig. 3a). As for the VL, we use ankle dorsiflexion torque and TA force synonymously, given the fixed-end isometric condition with the primary ankle dorsiflexor. We found a significant fixed effect of TA force on the change in S11 magnitude (p<0.001, Fig. 3b) and S11 phase (p<0.001, Fig. 3c). As anticipated, we found a significant fixed effect of TA force on activation (p<0.001, Fig. 3d) and fascicle length (p<0.001, Fig. 3e), with both remaining well-correlated with force (both $R^2$>0.95) during the isometric contractions.

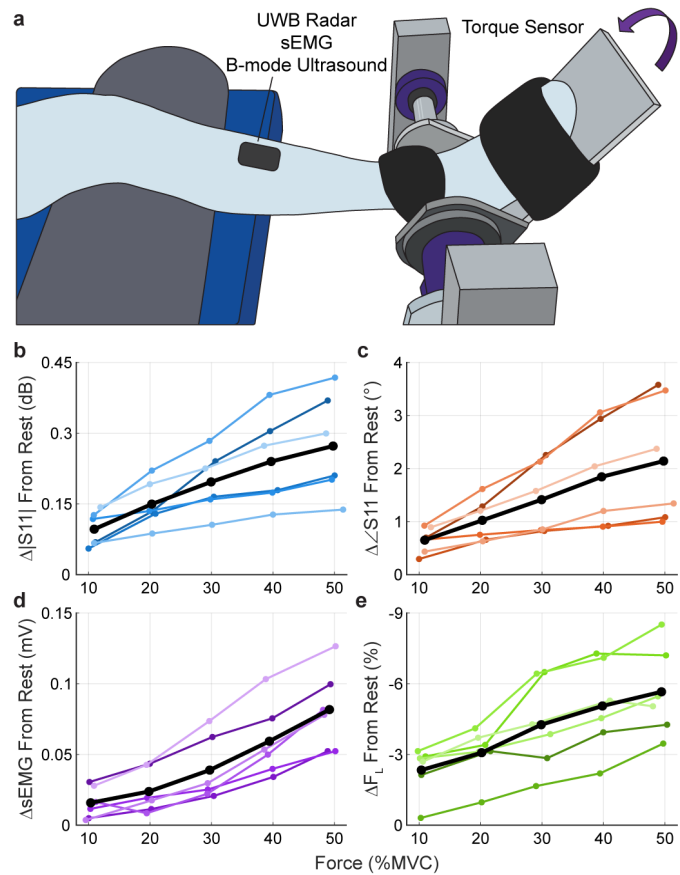

**Figure 3: TA forces during steady-state isometric ankle dorsiflexion contractions had a significant fixed effect on UWB radar S11 scans in both magnitude and phase.**
**a**, Illustration of isometric ankle dorsiflexion contractions experimental setup. **b**, Absolute change in S11 magnitude in the TA from rest, averaged across all frequencies (n = 6, p<0.001, LME). **c**, Absolute change in S11 phase in the TA from rest, averaged across all frequencies (n = 6, p<0.05, LME). **d**, Absolute change in sEMG magnitude in the TA from rest (n = 6, p<0.001, LME). **e**, Absolute change in fascicle strain in the TA from rest (n = 6, p<0.001, LME). For **b-e**, data is displayed for each participant (by lightness) with mean (black).

## Fatiguing Isometric Knee Extension Contractions

Next, we used the full UWB radar signal to estimate time-varying forces in the VL during isometric contractions under fatiguing conditions. In these experiments, muscle fatigue was purposefully introduced to decouple muscle activation from force output and determine the capabilities of UWB radar in the presence of fatigue (as a muscle fatigues, activation for a given force output increases). The time-varying estimates of force made use of the full UWB radar signal, utilising the magnitude and phase of each frequency of S11 and S21.

We performed a study on 18 participants during 20 minutes of intermittent isometric knee extension contractions to induce fatigue, with biofeedback consisting of trapezoidal force curves peaking between 8% and 45% MVC. We used a VNA and sEMG to record UWB radar scattering parameters (S11 and S21) and activation of the VL, with a load cell to measure knee extension torques (Fig. 4a). Again, during this isometric condition, we use knee extension torque and VL force synonymously. We successfully fatigued the VL muscle, as participants experienced an average 26±15% increase in activation for a given force at completion of the 20-minute contraction protocol (Fig. 4b). We observed a high inter-participant variability in the amount of fatigue experienced, with only one participant not experiencing significant fatigue (all other participants $p<0.001$). LSTM models (Fig. 4c) that estimate isometric forces in the VL from the UWB radar data (Extended Data Figure 1) were trained on individual participant data with a 60/20/20 training/validation/test split and evaluated with 5-fold cross-validation, achieving an average test $R^2$ of 0.988±0.005 and NRMSE of 2.7±0.6% (Fig. 4d-f, Extended Data Figure 2).

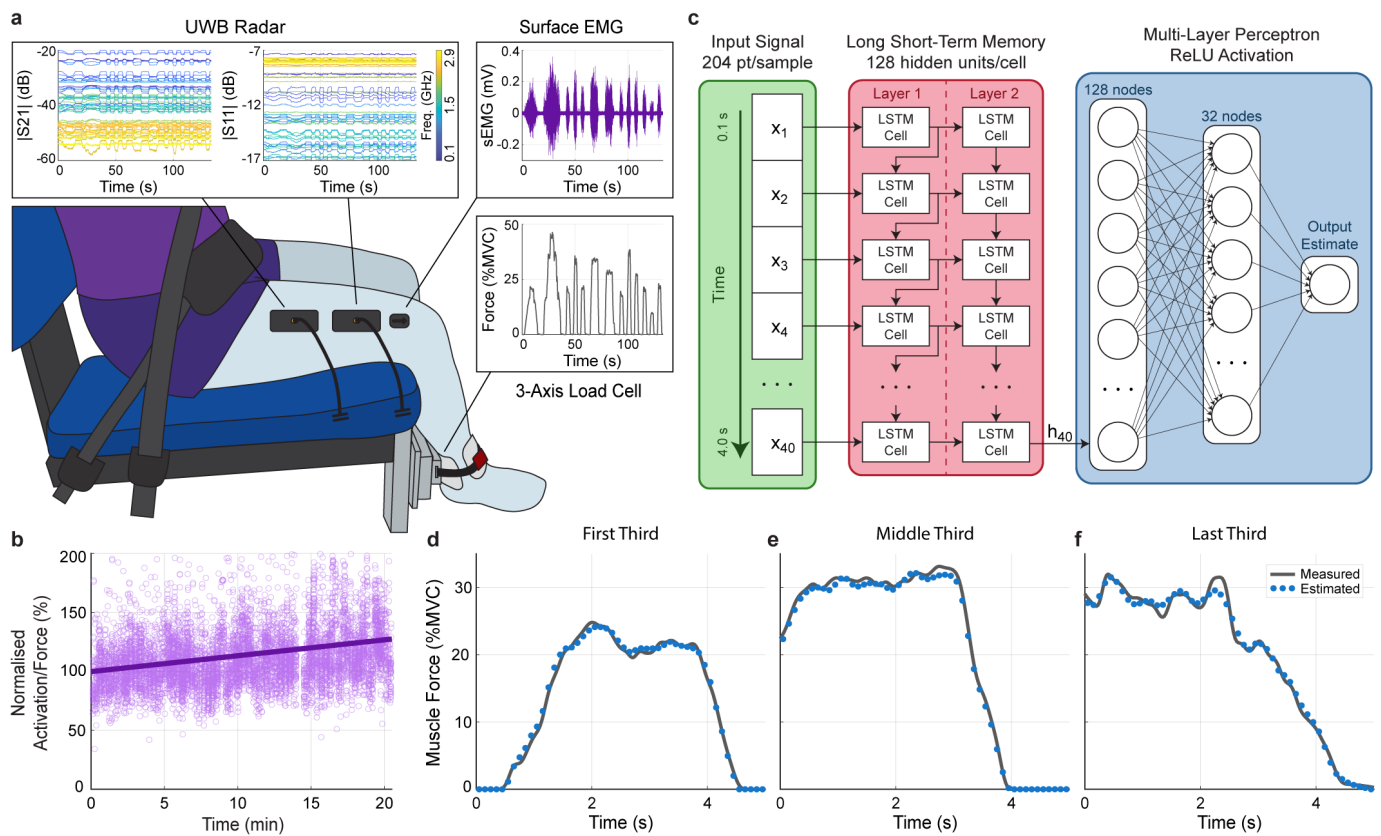

**Figure 4: Time-varying isometric muscle forces in the VL under fatiguing conditions were accurately estimated from UWB radar scans using long short-term memory machine-learning models.**
**a**, Illustration of fatiguing isometric knee extension contractions experimental setup. **b**, Representative plot showing a participant's ratio of activation to force over the duration of the 20-minute fatiguing protocol, with line of best fit (p<0.001, linear model). **c**, LSTM model architecture to estimate time-varying forces in the VL from UWB radar data. **d-f**, Representative test estimated force compared to measured force in the VL, taken from the (**d**) first, (**e**) middle, and (**f**) last third of the twenty-minute contraction protocol. **b** and **d-f** consist of data from the same participant, with median model test $R^2$. For **d-f**, both measured (grey line) and test estimated (blue dots) forces are displayed (all participants: $R^2$ = 0.988±0.005, NRMSE = 2.7±0.6%).

## Dynamic Knee Extension Contractions

During the non-fatiguing isometric contractions, activation and fascicle length remained well correlated with force, making it challenging to establish a causative link between muscle force and the change in UWB radar signal. To overcome this, we investigated the application of UWB radar to estimate time-varying VL forces during dynamic contractions, whereby muscle force is decoupled from both activation and fascicle length. By relying on carefully designed dynamic single-joint dynamometer experiments inspired from *ex vivo* muscle mechanics experiments [44, 45], we can decouple force from activation and fascicle length. While muscle force and joint torque remain closely linked, we cannot directly relate knee torque to VL force due to changes in moment arm that occur as knee angle varies [46]. Therefore, we refer to this measure in terms of torque.

We performed a study on 6 participants in an isokinetic dynamometer during a range of dynamic knee extension contractions at different speeds and torque levels (Fig. 5a). These included: passive movements at 30, 60, 90, and 120 °/s, providing fascicle length changes with minimal force and activation; isokinetic contractions at 30, 60, 90, and 120 °/s, giving concentric and eccentric contractions at fixed speeds; isokinetic contractions with a 45% MVC initial threshold at 30, 60, 90, and 120 °/s, providing different starting conditions to those without the threshold; and isotonic contractions at 15, 30, 45, and 60% MVC, giving concentric contractions at fixed torques. Together these different contraction types provide a broad sampling of different combinations of muscle force, activation, fascicle length, and fascicle velocity. During the dynamic contractions we used a VNA, sEMG and b-mode ultrasound to measure UWB radar scattering parameters (S11 and S21), activation, and fascicle dynamics (length, velocity, and pennation) of the VL, respectively (Fig. 5b), and an isokinetic dynamometer to measure knee extension torques and angles.

During the active isokinetic and isotonic conditions, activation and fascicle length were successfully decoupled from muscle force (both $R^2$<0.25, Extended Data Figure 3). We used linear models to estimate muscle forces from the UWB radar data, which were fit with an 80/20 training/test split and evaluated with 5-fold cross-validation. These models, when fit with individual participant data, were able to estimate dynamic muscle forces in the VL from the UWB radar data with an average test $R^2$ of 0.984±0.005 and NRMSE of 3.3±0.6% (Fig. 5c-e, Extended Data Figure 4).

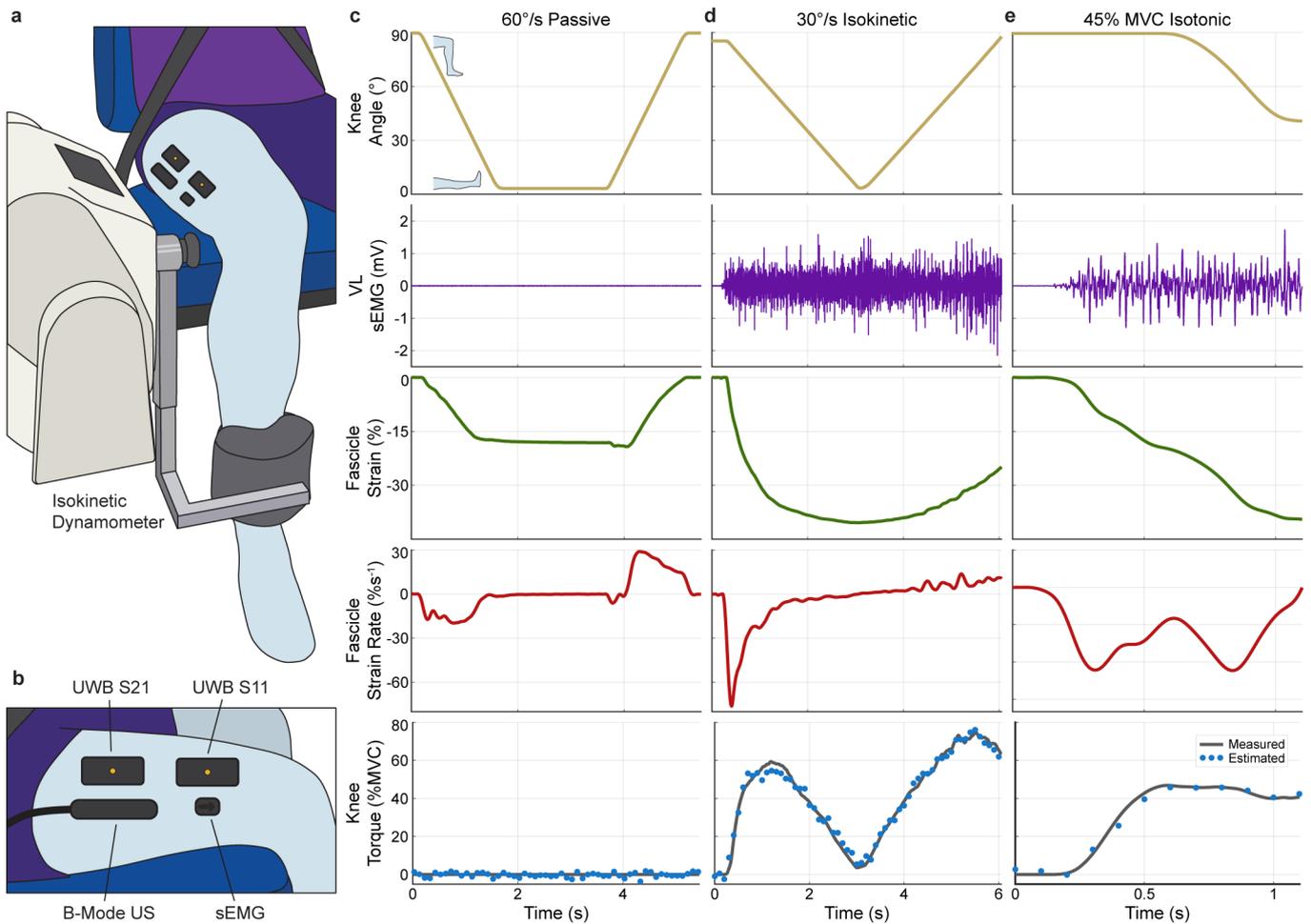

**Figure 5: Time-varying knee extension torques during dynamic conditions were accurately estimated from UWB radar scans of the VL using linear mixed-effects models.**
**a**, Illustration of dynamic knee extension contractions experimental setup. **b**, Placement of UWB radar antennas, sEMG, and B-mode ultrasound probe over the VL. **c-e** Representative plots of knee angle (yellow), VL sEMG (purple), fascicle strain (green), fascicle strain rate (red), and knee torque (grey and blue), during a (**c**) 60°/s passive, (**d**) 30°/s isokinetic, and (**e**) 45% MVC isotonic contraction. **c-e** consist of data from the same participant, with median model test $R^2$ and NRMSE performance metrics. For **c-e**, both measured (grey line) and test estimated (blue dots) torques are displayed on the bottom axes (all participants: $R^2$ = 0.984±0.005, NRMSE = 3.3±0.6%).

It is well established that the force generated by a muscle is dependent on activation, fascicle length, and fascicle velocity [39, 40] (Fig. 6a-d). We used linear models to determine how each of these parameters (as well as knee torque) affected the UWB radar data, across the frequency spectrum (100 MHz – 2.9G Hz) of both magnitude and phase of S11 and S21. Plotting the effect sizes of these parameters across the frequency spectrum (Fig. 6e-h) reveals the complex frequency-dependent nature of the UWB radar signal, and shows that each of these parameters, including torque, had frequency-dependent effects on the radar signal that were independent from each other. Pennation angle was also quantified, but given its correlations with fascicle length ($R^2$>0.75), only one architectural parameter could be used within this analysis. Using pennation angle rather than fascicle length results in very similar parameter effect sizes (Extended Data Figure 5).

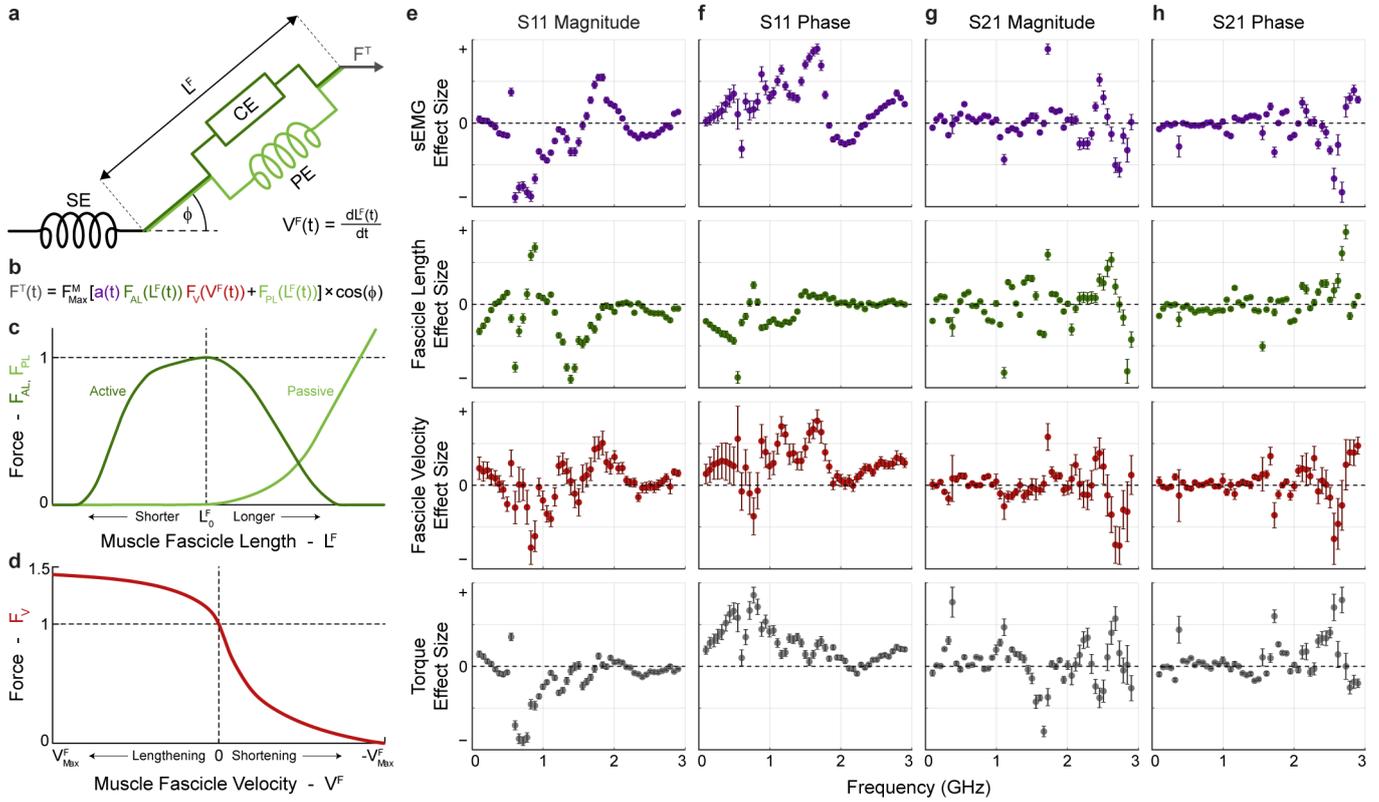

**Figure 6: Multiple factors known to influence muscle force had different frequency-dependent effects on the UWB radar signal.**
**a**, Representation of a Hill-type muscle/tendon model consisting of an active contractile element (CE), parallel elastic element (PE), and series elastic element (SE). **b**, Typical equation for a hill-type muscle model, with output force a function of activation ($a(t)$), fascicle length ($L^F(t)$), fascicle velocity ($V^F(t)$), and pennation angle ($\phi$). **c**, Typical force-length relationship of a muscle fascicle, with both active ($F_{AL}$, dark green) and passive ($F_{PL}$, light green) components. **d**, Typical force-velocity relationship of a muscle fascicle ($F_V$). **e-h**, Relative effect sizes of sEMG, fascicle length, fascicle velocity, and torque on the UWB radar signal (**e**) S11 magnitude, (**f**) S11 phase, (**g**) S21 magnitude, and (**h**) S21 phase, across the 100 MHz – 2.9 GHz frequency range. For **e-h**, error bars represent the 95% confidence interval, with data from all participants (n = 6) across all contraction conditions.

## Discussion

In this work, we completed a series of experiments to demonstrate the potential of UWB radar as an innovative tool to non-invasively estimate *in vivo* musculoskeletal forces – across muscle architectures, static and dynamic contractile conditions, and in the presence of fatigue. First, we established ultra-wideband radar scans of muscle can reliably track isometric forces in a unipennate VL and a bipennate TA. Following this, LSTM models using UWB radar data estimated time-varying forces in the VL with high accuracy during a series of fatiguing isometric knee extension contractions, situations where muscle activation becomes a poor predictor of force. Next, we demonstrated that linear models of radar signals can predict knee torque with high accuracy during a range of active and passive dynamic contractions, where forces are decoupled from activation and fascicle length. These estimates of musculoskeletal forces across a wide range of conditions highlight UWB radar's potential applications for clinicians, biomechanists, and engineers as a tool that gives valuable insight into the muscles that power movement.

The musculoskeletal force estimates provided by UWB radar offer several advantages over existing technologies, with accuracy standing out as most critical. We obtained VL force estimates with exceptional accuracies: test $R^2$ of 0.988 and NRMSE of 2.7% during isometric fatiguing contractions, and knee torque estimates with test $R^2$ of 0.984 and NRMSE of 3.3% during dynamic contractions. These accuracies were achieved with rigorous approaches to prevent model over-fitting, including a training/test data split and 5-fold cross-validation. Compared with prior approaches, UWB radar demonstrates a clear step forward: using A-mode ultrasound to estimate elbow and knee torques during isokinetic contractions with individual models reported $R^2$ of 0.92 [47], while multi-frequency electrical impedance myography (EIM) to estimate elbow torques during isometric, isotonic, and damped contractions achieved $R^2$ of 0.79, 0.64, and 0.94, respectively [32]. Another EIM study reported $R^2$ of 0.83 during isometric elbow flexion contractions using a multi-participant model and leave-one-out cross-validation [31]. Importantly, we show that UWB radar maintains high accuracy across conditions that often challenge previous methods, including isometric and dynamic tasks, non-fatiguing and fatiguing states, and across both ankle and knee joints. Further, EMG-based approaches typically fail during dynamic conditions and under fatigue [5], and other technologies have only been evaluated for force estimation in non-fatigued states [10, 31, 47]. While high accuracy estimates of joint torque have been reported using tendon tensiometry ($R^2$ of 0.96 and 0.98 from the Achilles and patellar tendon, respectively [15]), its utility is limited to superficial tendons, and accuracy varies with tendon morphology and loading conditions [15, 16, 17].

Our experimental design was structured to disentangle the key parameters that influence muscle force in a Hill-type model – activation, fascicle length, and fascicle velocity. In the fatiguing protocol, activation was progressively decoupled from VL force, as greater activation was required to sustain a given force level over the 20-minute protocol. The dynamic contractions further uncoupled mechanical and physiological factors by disassociating activation, fascicle length, and fascicle velocity from muscle force (Extended Data Figure 3) – enabling our capacity to demonstrate the UWB radar signal was not simply responding to a single factor such as fascicle length, which remains strongly correlated with force during isometric contractions. Specifically, the passive contractions altered fascicle length and velocity with minimal activation and torque; the isokinetic contractions altered fascicle length and velocity with near-maximal activation; and the isotonic contractions set fixed torques across varying activations, fascicle lengths, and velocities. Across these varied conditions, UWB radar consistently provided high accuracy estimates of knee torque, demonstrating its capacity to estimate forces that are independent of activation, fascicle length, and fascicle velocity, as is commonly the case during real-world movements. Moreover, this approach enabled us to probe the origins of the radar signal itself, revealing complex, frequency-dependent interactions with activation, fascicle length, fascicle velocity, and torque – providing important insights into the physiological and mechanical basis of what the UWB radar is measuring.

Fundamentally, the UWB radar signal responds to changes in the electromagnetic properties of the contracting muscle, namely the conductivity and permittivity. We have shown that these are associated with changes in activation, fascicle length, fascicle velocity, and torque. However, this does not explain the physiological mechanisms underpinning changes in the radar signal. The effect of activation may be explained by $Ca^{2+}$ release from the sarcoplasmic reticulum during muscle activation [48, 49], where ions transition from a state bound to the negatively charged calsequestrin to free positive ions, potentially increasing conductivity and reducing permittivity. The effect of fascicle length may be explained by its correlation to fascicle pennation angle, and the known difference in EM properties when measured perpendicular to muscle fibres compared to parallel [41]. Advancing our understanding of the physiological and mechanical features underpinning changes in UWB radar signals will likely require a combination of *ex vivo* experiments to directly quantify tissue-level parameters and simulations to test mechanistic hypotheses. These insights will guide the optimization of antenna design and selection of frequency bands tailored to specific muscles or tasks, ultimately improving the utility of UWB radar for musculoskeletal applications.

In this study, we used different modelling approaches to estimate musculoskeletal forces during fatiguing isometric contractions compared to dynamic contractions, reflecting varying amounts of available data. As expected, more complex LSTM models (Extended Data Figure 2) slightly outperformed linear models (Extended Data Figure 4) in estimating forces. However, the high accuracy of linear models highlights that simpler models may be sufficient for many applications. All models were trained and tested on an individual participant basis, given the substantial inter-participant variability in the UWB radar data. Future work will focus on strategies to normalize radar data between participants for general models, which can be trained and tested on different participants. Alternatively, if inter-participant variations remain high and blind normalisation between participants is impractical, a calibration re-training step utilising a set of pre-established contractions may be required.

The potential applications of UWB radar technology to non-invasively estimate musculoskeletal forces are vast: as a real-time control signal for exoskeletons and prosthetics that enhance or restore motor performance; for prevention and detection of injury in sports; or for real-time control of functional electrical simulation-based rehabilitative devices capable of counteracting muscle fatigue. However, there remain some technical challenges to overcome. The coaxial cables that connect the antennas and VNA are sensitive to movement and need to be maintained at short lengths. Adding ferrite cores along the cables results in significant improvements, but at the cost of increased cable weight. The antennas used here, while reasonably small (30×70mm), are still somewhat bulky, and may not be suitable for small muscles. Designing smaller antennas may require higher frequencies. Lastly, our UWB radar scans consist of 51 frequencies, providing a sampling rate of ~10 Hz with our VNA. While frequency analysis confirmed 10 Hz to be sufficient for all of the data captured here, a higher sampling rate may be beneficial for other tasks, such as walking. This can be achieved in future research with either a faster VNA, or by sampling at fewer individual radar frequencies.

In conclusion, we demonstrate the potential for UWB radar to non-invasively estimate *in vivo* musculoskeletal forces. This has been established in lower limb muscles with varying architectures, and across a range of contraction conditions, including isometric, fatiguing, passive, and dynamic. Through experimentally uncoupling neuromuscular, structural, and physiological factors, we provide mechanistic insights into the relationship between force and UWB radar. Looking forward, UWB radar has the potential to transform the study and application of human movement, enabling real-time monitoring of musculoskeletal loads during functional tasks, guiding the design of wearable assistive devices, and opening new avenues for personalized rehabilitation and performance optimization.

# Methods

### Experimental Overview

Four experiments were conducted to explore the use of UWB radar as a tool to estimate muscle forces, across different muscles (VL and TA) and contraction conditions (steady-state/time-varying; fatiguing/non-fatiguing; isometric/dynamic; active/passive). We began investigating how the UWB radar signals change under simple, steady-state isometric conditions in the unipennate VL. This was subsequently repeated in the bipennate TA. Next, we looked to use this signal to estimate time-varying muscle forces during isometric contractions under fatiguing conditions, where a muscle's activation ceases to be a good estimator of force. Finally, participants completed a range of dynamic knee extensions contractions (passive, isokinetic, isotonic), to estimate knee torques that are decoupled from activation and fascicle length.

For all sensors and conditions, b-mode ultrasound was first used to characterise muscle boundaries to ensure accurate placement of sensors over the muscle of interest. In all cases where sEMG was collected from the VL, it was additionally simultaneously recorded from the vastus medialis (VM) to monitor changes in load-sharing between VL and VM. For all conditions, maximum voluntary contractions (MVCs) were found at the beginning of data collections. For the quadriceps, these were measured at a knee flexion angle of 70°, and for the TA with an ankle angle of 110°. Participants performed MVCs with verbal encouragement and real-time torque biofeedback for a minimum of three attempts, or until the maximum torque produced in an attempt was less than their overall maximum. Participants received two minutes rest between each MVC attempt.

Ultra-wideband (UWB) radar data were collected using a vector network analyser (VNA) (LiteVNA 64, Zeenko, China), and either one or two body-coupled antennas [43]. Coaxial cables connected the antennas to the VNA, with ferrite cores along the length of the cables to prevent the inner metallic shield from acting as an antenna. In the case of one antenna being used, the scattering parameter S11 (input reflection coefficient) was collected, while with two antennas the scattering parameters S11 and S21 (forward transmission coefficient) were collected. These scattering parameters are complex, with both magnitude and phase. Scattering parameters were collected at approximately 10.8 Hz, sampling 51 evenly spaced frequencies from 100 MHz to 2.9 GHz.

**Participants**

Sixteen healthy adults (9 females and 7 males; age = 22.3±3.7 years; height = 1.72±0.09 m; mass = 65.5±9.2 kg) participated in the isometric knee extension tests. While twenty participants were recruited, data from four were excluded due to poor ultrasound quality. Six healthy adults (3 females and 3 males; age = 24.7±1.8 years; height = 1.75±0.12 m; mass = 71.5±14.1 kg) participated in the isometric ankle dorsiflexion test. Eighteen healthy adults (9 females and 9 males; age = 22.4±3.5 years; height = 1.72±0.09 m; mass = 65.3±9.6 kg) participated in the fatiguing isometric knee extension tests. While twenty participants were recruited, one participant was excluded due to their MVC exceeding the measurement limits of our force sensor, and data from another was excluded due to poor quality of the sEMG signal. Six healthy adults (3 females and 3 males; age = 24.5±2.2 years; height = 1.75±0.11 m; mass = 74.4±10.4 kg) participated in the dynamic knee extension test. While eight participants were recruited, two were excluded due to insufficient space on the skin over their VL for all of the required sensors. All participants provided written informed consent before participation, and all experimental protocols were approved by the Human Research Ethics Committee of The University of Queensland (2023/HE000932).

**Isometric Knee Extension Contractions**

Sixteen participants completed the isometric knee extension test. Participants completed isometric knee extension contractions with their right leg and were provided biofeedback via a monitor at five levels of normalised torque (10, 20, 30, 40, and 50% MVC), with their knee flexed to 70°. Contractions consisted of a one second ramp up, three second hold, and one second ramp down, with a five second rest between each contraction. Participants were given several attempts at the task to familiarize themselves with the setup, and in the case of multiple successful attempts, a participant's results were taken as the mean change in signal across attempts.

Knee extensions torque was found with a custom acquisition board and 3-axis load cell (F3F-500-500-1k, Forsentek, Japan) connected to the ankle (40 Hz) as the magnitude of the 3-axis change from rest for each contraction, normalised to MVC. UWB radar scattering parameter S11 from a body-coupled antenna over the VL (placed centrally over the muscle belly) was recorded (10.8 Hz; 51 frequencies from 100 MHz – 2.9 GHz). B-mode ultrasound (MicrUs EXT-1H, Telemed, Lithuania; and LV8-4L65S-3, Telemed, Lithuania) of the VL (transducer placed proximally to the UWB antenna) captured changes in fascicle length, which were quantified using UltraTrack (ver. 5.0) [50]. sEMG (Trigno Avanti, Delsys, USA) data from the VL (placed distally to the UWB antenna, 2048 Hz) was band-pass filtered (10Hz – 500 Hz), notch filtered (50 Hz, and harmonics as necessary), rectified, and low-pass filtered with a zero-lag $4^{th}$ order Butterworth filter with cut-off at 5 Hz. The change from rest to contracted for each measure (S11 magnitude and phase, fascicle length, and sEMG magnitude) at the plateau of each torque level was determined. The measures of S11 magnitude and phase change were taken as the mean absolute change across the frequency range.

Linear mixed-effect (LME) models were used to determine the effect of isometric muscle forces on the change from rest to contracted in S11 magnitude, S11 phase, sEMG magnitude, and fascicle length of the VL. An LME was created for each measured parameter in each muscle, with the measured parameter as response variable, and isometric muscle force as predictor variable with a random independent intercept and slope that varied by participant. Fixed effects were analysed with Bonferroni corrections, and the level of statistical significance was set at $p<0.05$.

**Isometric Ankle Dorsiflexion Contractions**

Six participants completed isometric ankle dorsiflexion contractions with their right leg following a biofeedback ankle torque trace at five different levels of normalised torque (10, 20, 30, 40, and 50% MVC), with an ankle angle of 110°. Contractions consisted of a one second ramp up, three second hold, and one second ramp down, with a five second rest between each contraction. Participants were given several attempts at the task to familiarize themselves with the setup, and in the case of multiple successful attempts, a participant's results were taken as the mean change in signal across attempts.

Ankle dorsiflexion torque was found with a torque transducer (TRE-50K, DACELL, South Korea), and was normalised to MVC. Due to the limited space over the TA muscle belly, these contractions were completed separately for each of the UWB S11, sEMG, and b-mode ultrasound (ArtUs EXT-1H, Telemed, Lithuania; and LV8-5N60-A2, Telemed, Lithuania) measures, which were recorded and processed consistent with the isometric knee extension test (with each sensor placed centred over the TA muscle belly). The change from rest to contracted for each measure (S11 magnitude and phase, fascicle length, and sEMG magnitude) at the plateau of each torque level was determined. The measures of S11 magnitude and phase change were taken as the mean absolute change across the frequency range.

Linear mixed-effect (LME) models were used to determine the effect of isometric muscle forces on the change from rest to contracted in S11 magnitude, S11 phase, sEMG magnitude, and fascicle length of the TA as they were for the isometric knee extension contractions.

**Fatiguing Isometric Knee Extension Contractions**

Eighteen participants completed a fatiguing isometric knee extension protocol, involving twenty minutes of intermittent isometric knee extension contractions with their right leg. Participants were given biofeedback of torque consisting of trapezoidal force curves peaking between 8% and 45% MVC, divided into a fourteen-minute and a six-minute section with a thirty-second break between sections. The purpose of the thirty-second break part-way through the contractions was to allow some recovery from muscle fatigue, which aimed to increasing the non-linearity of the fatigue response.

sEMG of the VL and knee extension torques were recorded and processed consistent with the isometric knee extension test. Two body-coupled antennas were placed over the VL (laterally centred, distally to each other), and scattering parameters S11 and S21 were recorded (10.8 Hz; 51 frequencies from 100 MHz – 2.9 GHz).

UWB radar scans, sEMG data , and knee extension torques were subsequently resampled to a synchronous 10 Hz. Drift in the resting-point torque that occurred throughout the data collection was corrected for, and perturbations in the signal from participant movements during rest were manually discarded.

A participant's muscle fatigue was quantified over the duration of the data collection as a linear model fitted to the ratio of the VL's sEMG magnitude to force against time (with higher activation for a given force indicating higher muscle fatigue), at time points where both signals were at least 5% of their maximum. The statistical significance of this effect of time on the ratio was analysed with Bonferroni corrections, and the level of statistical significance was set at $p<0.05$. A participant's fatigue metric, where a higher fatigue metric indicates a participant experienced more fatigue, was then taken as the value of this fit at the end divided by the start of the data collection.

LSTM models were trained on individual participant data to estimate time-varying isometric muscle force from the UWB radar data (magnitude and phase of both S11 and S21). To ensure that various levels of muscle fatigue were uniformly represented in each of the training, validation, and test sets, the data was first segmented into 30-second blocks, with each block then split into training (60%), validation (20%), and test (20%) sets. An augmented copy of the training data was created by interpolating at the same sample rate, offset by half a sample period, doubling the size of the training set. The model architecture consisted of a two-layer LSTM, with 40 cells per layer (4-seconds at 10 Hz) and 128 hidden units per cell. The final cell's hidden state became the input of a multi-layer perceptron (size 128/32/1) with ReLU activation. Models were evaluated using 5-fold cross-validation, with normalised RMSE (NRMSE) and coefficient of determination ($R^2$) as performance metrics.

**Dynamic Knee Extension Contractions**

Six participants completed sixteen sets of right knee extension contractions in an isokinetic dynamometer (System 4 Pro, Biodex, USA) under a series of different contractile conditions: passive (30, 60, 90, 120° $s^{-1}$), isokinetic (30, 60, 90, 120° $s^{-1}$), isokinetic with 45% MVC initial threshold (30, 60, 90, 120° $s^{-1}$), and isotonic (15, 30, 45, 60% MVC). Range of motion was set to 0° to 90° for all contractions, with 0° defined as full knee extension. Participants were verbally instructed to fully relax during the passive motion, exert near-maximum effort during both the concentric and eccentric portions of the isokinetic VL contractions, and extend as fast as possible during the concentric portion of the isotonic VL contractions. Participants completed minimum three repetitions of each condition, of which one was used for analysis. The order of contractions was pseudo-randomised, and participants received three minutes of rest between each condition to mitigate fatigue and were allowed additional rest as needed.

Knee torque, angle, and angular velocity were recorded by the dynamometer (500 Hz), with torque values being automatically gravity corrected and subsequently low-pass filtered with cut-off at 5 Hz. During passive conditions, knee torques were defined as zero. Two body-coupled antennas were placed medially over the VL (distally to each other), and scattering parameters S11 and S21 were recorded (10.8 Hz; 51 frequencies from 100 MHz – 2.9 GHz). sEMG was recorded from the VL (placed laterally to the distal UWB antenna, 2048 Hz), and was processed as it was for the isometric knee extension test and normalised to MVC activation. B-mode ultrasound (ArtUs EXT-1H, Telemed, Lithuania; and LV8-5N60-A2, Telemed, Lithuania) video of the VL (transducer placed laterally to the proximal UWB antenna, 100 Hz) captured changes in fascicle length, pennation, and velocity, which were quantified using UltraTrack (ver. 5.0) [50], low-pass filtered with cut-off at 5 Hz, and normalised to values at rest. All data were subsequently resampled to a synchronous 10 Hz. If a participant completed an isotonic condition in under 300 ms (i.e., fewer than three sample points), this condition was not considered in the analysis for this participant.

LME models were used to determine the correlation of knee torque with fascicle length and sEMG magnitude during the active conditions of the dynamic knee extension test. These models were consistent with those used for the isometric knee extension test, and had either fascicle length or sEMG magnitude as the response variable, and torque as the predictor variable with a random independent intercept and slope that varied by participant.

Next, LME models were used to estimate time-varying dynamic knee torques from the UWB radar data (magnitude and phase of both S11 and S21). The models started with torque as the response variable, and all 204 UWB radar traces as predictor variables, with a random intercept that varied by condition with each participant. Predictor variables were iteratively removed by the largest p-value, until all predictors had a significant fixed effect ($p<0.001$) and there were no more than fifty predictor variables. Models were evaluated with 5-fold cross-validation, where each fold's test set consisted of every fifth data point (with an offset between folds) while models were fit to the remaining 80% of the data. Normalised RMSE (NRMSE) and coefficient of determination ($R^2$) of the test set estimation were used as model performance metrics.

Finally, LME models with data from all participants and contractile conditions were used to determine the effect of different parameters, namely sEMG magnitude, fascicle length, fascicle velocity, and knee torque, on the UWB radar signal at different frequencies. A linear model was created for each of the magnitude and phase of S11 and S21 for each of the 51 sample frequencies (204 linear models in total), with the UWB trace as response variable, and each of sEMG magnitude, fascicle length, fascicle velocity, and torque as predictor variables.

# Acknowledgements

We would like to acknowledge and thank Prof. Amin Abbosh and Dr. Hadi Mousavi from The University of Queensland for their advice regarding the body-coupled UWB antennas.

# Extended data figures

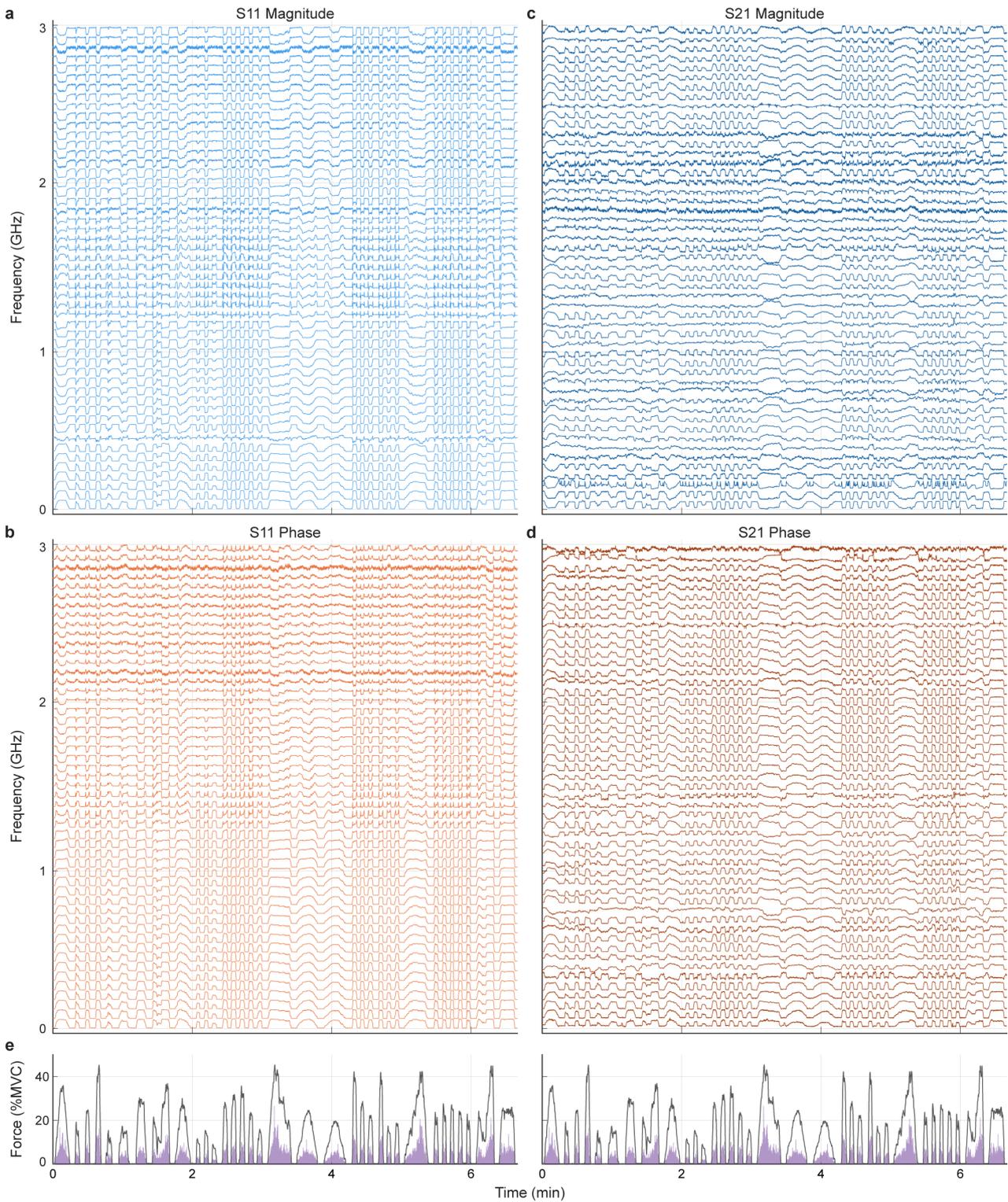

**Extended Data Figure 1: Full UWB radar signal (magnitude and phase of S11 and S21) aligned to VL force and activation during isometric knee extension contractions.**
**a-d**, All UWB signal traces from a series of isometric knee extension contractions, individually normalised, and vertically positioned by frequency, for (**a**) S11 magnitude, (**b**) S11 phase, (**c**) S21 magnitude, and (**d**) S21 phase. **e**, Torque (grey) and activation (purple, arbitrary units) traces corresponding to **a-d**. For **e**, the traces are repeated (left and right) to be aligned with **a-d**.

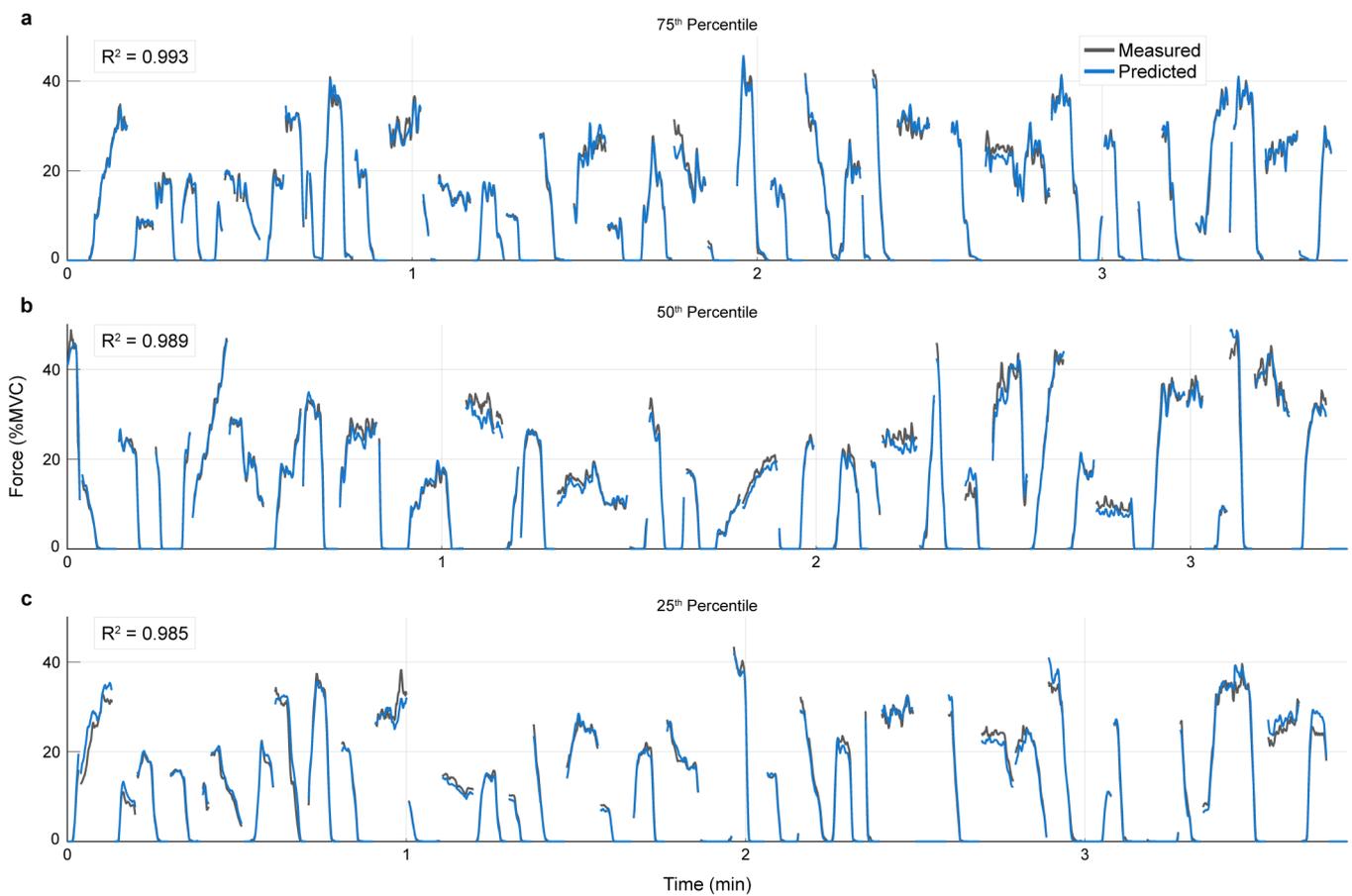

**Extended Data Figure 2: Additional long short-term memory machine-learning model estimates of isometric VL force under fatiguing conditions.**
**a-c**, Plots showing measured (grey) and LSTM test estimated (blue) isometric VL force, each for one-fold of a participant's data, at the (**a**) 75th, (**b**) 50th (i.e., median), and (**c**) 25th percentile of $R^2$, with $R^2$ values of 0.993, 0.989, and 0.985 respectively. For **a-c**, due to the nature of the data split, these consist of discontiguous segments from across the complete 20-minute fatiguing protocol, with the relative order of segments maintained.

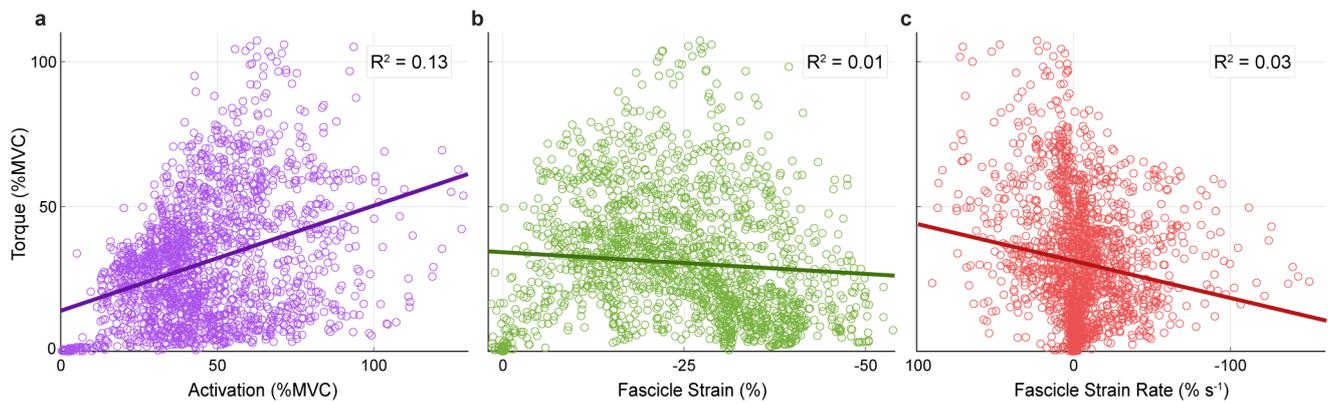

**Extended Data Figure 3: Activation, fascicle length, and fascicle velocity in the VL were successfully decoupled from knee torque during the active dynamic contraction conditions.**
**a**, Knee torque against activation in the VL for all active conditions in all participants (n = 6), with line of best fit ($R^2 = 0.13$). **b**, Knee torque against fascicle strain in the VL for all active conditions in all participants (n = 6), with line of best fit ($R^2 = 0.01$). **c**, Knee torque against fascicle strain rate in the VL for all active conditions in all participants (n = 6), with line of best fit ($R^2 = 0.03$). For **a-c**, LME models accounting for interparticipant variability (random intercept and slope by participant) result in an improved fit ($R^2 = 0.20$, $R^2 = 0.21$, and $R^2 = 0.21$, respectively), though all are still well decoupled from force.

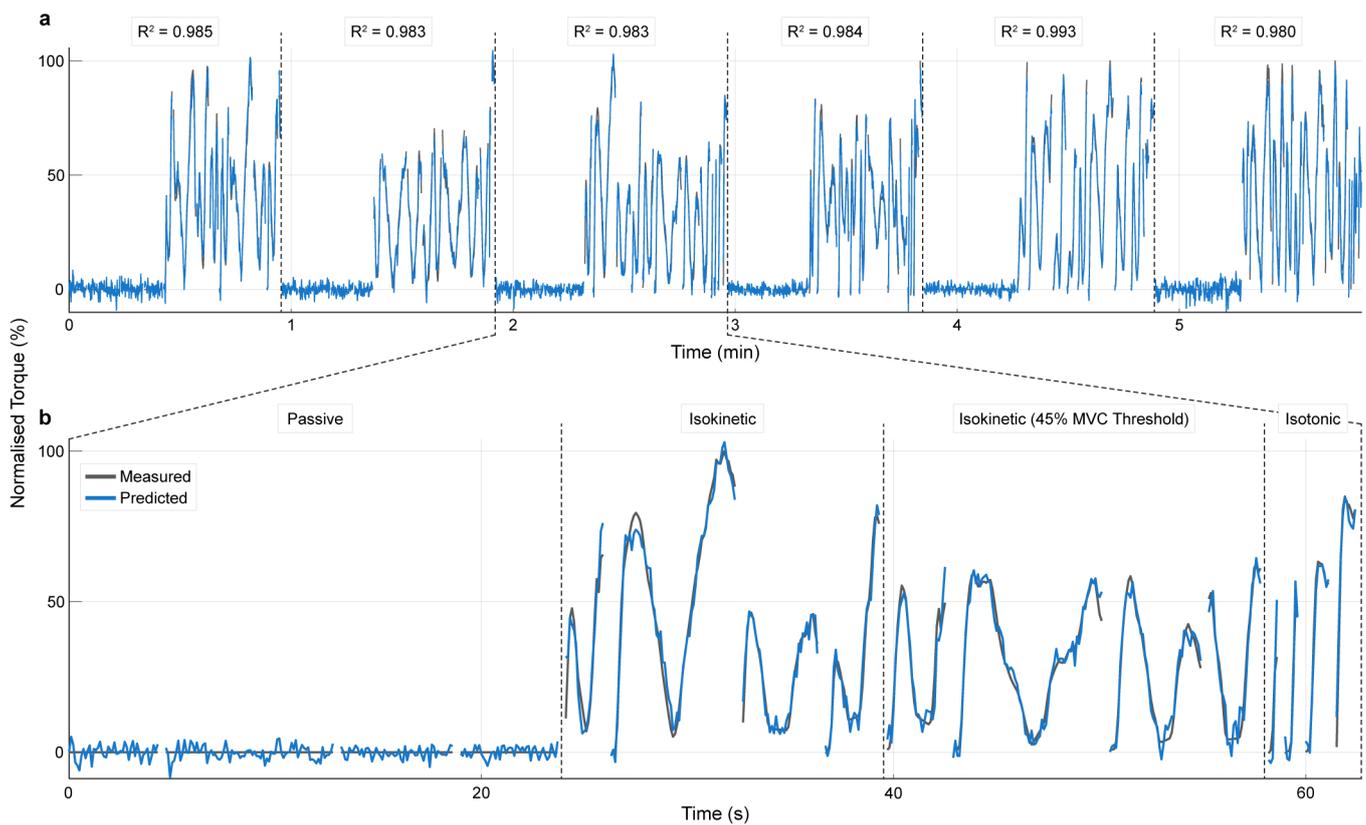

**Extended Data Figure 4: All linear model estimates of knee torque during dynamic contractions.**
**a** and **b**, Plots showing measured (grey) and linear model test estimated (blue) normalised dynamic knee torques, (**a**) for all six participants (split by participant), and (**b**) for the participant with median test performance (split by contraction type). For **a** and **b**, these consist of discontiguous segments for each contraction condition, in order: passive (120°/s, 30°/s, 60°/s, 90°/s), isokinetic (120°/s, 30°/s, 60°/s, 90°/s), isokinetic with 45% MVC threshold (120°/s, 30°/s, 60°/s, 90°/s), and isotonic (15% MVC, 30% MVC, 45% MVC, 60% MVC).

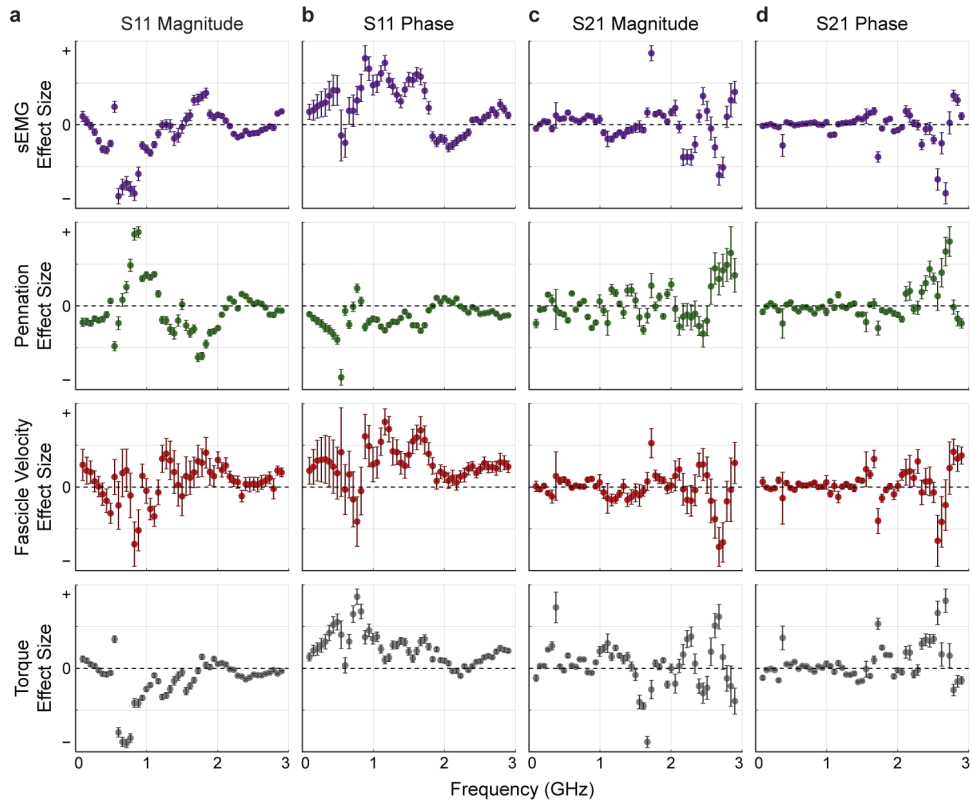

**Extended Data Figure 5: Using pennation instead of fascicle length in the LME models results in similar effect sizes across the frequency range.**
**a-d**, Relative effect sizes of sEMG, pennation, fascicle velocity, and torque on the UWB radar signal (**a**) S11 magnitude, (**b**) S11 phase, (**c**) S21 magnitude, and (**d**) S21 phase, across the 100 MHz – 2.9 GHz frequency range. For **a-d**, error bars represent the 95% confidence interval, with data from all participants (n = 6) across all contraction conditions. **a-d** correspond to Fig. 6e-h, with the models using pennation instead of fascicle length.